\documentclass[letterpaper]{article}

\usepackage[T1]{fontenc}

\usepackage{geometry}
\usepackage{setspace}

\usepackage[style=chem-acs,doi=true,articletitle=true]{biblatex}
\usepackage{graphicx}
\usepackage{float}
\newfloat{scheme}{htbp}{los}
\floatname{scheme}{Scheme}
\floatname{chart}{Chart}
\newfloat{graph}{htbp}{loh}

\usepackage{chemformula} 
\usepackage[version = 4]{mhchem} 

\newcommand{\be}{\begin{equation}}
\newcommand{\ee}{\end{equation}}

\newcommand{\Fig}[1]{Fig.~\ref{fig:#1}}

\newcommand{\Tab}[1]{Table~\ref{tab:#1}}
\newcommand{\Tabs}[2]{Tables~\ref{tab:#1} and ~\ref{tab:#2} }

\newlength{\wholefigwidth}
\newlength{\halffigwidth}
\usepackage{authblk}
\author[1,2]{Grace M. Lu}
\author[1]{Dallas R. Trinkle*}
\affil[1]{Department of Materials Science and Engineering, University of Illinois at Urbana-Champaign, Urbana, Illinois 61801, USA}
\affil[2]{Department of Mechanical and Aerospace Engineering, University of California at Irvine, Irvine, California 92697, USA}

\title{Thermodynamics and kinetics of lithium at the silver-lithium battery interface}
\date{*Email: dtrinkle@illinois.edu}

\begin{document}

\maketitle

\begin{abstract}
 Silver interlayers have been shown to enable smooth lithium deposition and cycling in anode-free solid-state batteries. Here, we report the atomic structure of the Ag and Li interface, showing that Li preferentially plates as FCC on both the $(111)$ and $(100)$ Ag surfaces. This forms an energetically favorable coherent interface with Ag, while the BCC phase forms a semi-coherent interface due to large lattice mismatch. We also calculate vacancy formation energies and migration energies for Li diffusion through the interface. We show that vacancy formation energies increase at the interface, leading to an energetic driving force for vacancies to diffuse away from the interface. Additionally, the migration barriers for vacancies from the Ag to the Li are small (29 meV), and therefore promote rapid alloying between Ag and Li. Rapid Li diffusion kinetics directly at the interface leads to smooth deposition of Li, reducing the onset of dendrites. However, diffusion in the 2nd and 3rd Li layers is slower compared to bulk FCC or BCC Li, leading to kinetically hindered alloying when multiple layers of pure Li form. The diffusion kinetics for Ag nanoparticles may be improved by alloying with Mg to expand the Ag lattice constant while forming a solid solution with both Ag and Li.
\end{abstract}


\vspace{12pt}

Improving the performance of current lithium-ion batteries for applications that demand high energy densities, such as electrical vehicles, is an important open problem. Anode-free lithium batteries, in which no anode or excess lithium metal is included, have a theoretical capacity of more than ten times that of graphite (3820 mAh cm$^{-1}$) \cite{tarascon2001issues}. These anode-free lithium batteries rely on reversible plating and stripping of Li metal onto a current collector \cite{rosso2001onset}. Unfortunately, uncontrollable dendrite growth at the interface between the electrolyte and electrode in batteries leads to internal short circuits \cite{feng2020mitigating}. The key to suppressing dendrite growth is maintaining a smooth Li surface through deposition and stripping at the Li metal-electrolyte interface \cite{lu2014stable, hao2019electrodeposition, jackle2014microscopic}. Unfortunately, common solid electrolytes can have surface contaminants made of highly lithiophobic compounds, such as \ce{Li2Co3} and \ce{Li2S} \cite{gao2020surface, sharafi2017surface}, which make for rough interfaces, contact loss, increased interfacial resistance, and the trapping of vacancies at the interface forming voids \cite{yang2021maintaining}. These voids have been tied directly to dendrite formation as they create hotspots with localized pressures and overpotentials \cite{kasemchainan2019critical}. Recent promising work has shown that a solid electrolyte battery containing an interlayer, which consists of a 3:1 ratio of Ag nanoparticles and carbon black between a solid sulfide electrolyte \ce{Li6PS5Cl}, can lead to a longer cycle life (\textgreater 1000 cycles) \cite{samsung2020, samsung2023}, which opens the door to commercial adoption of anode-free lithium batteries.

The success of the Ag nanoparticle interlayer in anode-free lithium batteries is the improvement in lithium deposition, seemingly driven by alloying between the Ag and Li. Using SEM, Lee et al.\ found smooth deposition of Li, no evidence of Li dendrites forming after 100 cycles, and no residual Li deposits. They theorized that the Ag reduces the conductivity and the nucleation energy of Li metal, allowing for smoother deposition. The carbon serves as a separator that keeps the Li from direct contact with the electrolyte and improves Li transport, allowing for rapid transport of Li through the interlayer \cite{samsung2023}. 
The presence of $\gamma_2$-\ce{Li9Ag4} at the boundary signifies an alloying mechanism between the Ag and Li, which is vital to explain the performance. The formation of these Li-Ag mixed phases throughout charging and discharging has been studied experimentally through XRD \cite{magdysyuk2023, ko2025mechanism}. Specter-Jolly et al.\ found that at rates of charging of 2 mA cm$^{-2}$ or lower, after the carbon is fully lithiated to form LiC$_6$, Li$_{x}$Ag solid solutions form \cite{magdysyuk2023}. New intermetallic phases, like $B_2$ LiAg, $\gamma_2$ Li$_9$Ag$_4$, and $\gamma$-3 Li$_{10}$Ag$_3$, appear in order of increasing Li content. While the XRD measurements cannot directly characterize the Li metal because Li is a weak X-ray scatterer, the quantity of charge exceeds the capacity of the LiC$_6$ and Li$_{10}$Ag$_3$, so Li metal is believed to form. 
From a computational perspective, Xie et al.\ fit cluster expansions for BCC, FCC, and $\gamma$-structured Ag-Li \cite{xie2024microscopic} using DFT. They concluded that at 300 K, when the atomic percent of Li is greater than 46.6\%, the BCC solid solution is stabilized by entropy and is more energetically favorable than any of the $\gamma$ phases. Thus, they theorized that the Li-rich Ag-Li solid solution retains a driving force to absorb Li, allowing for smooth insertion of Li. While little is known about the atomic mechanism, replacing Ag with Mg, which also forms a solid solution with Li  \cite{pasta2023}, does not result in similar performance.

Even though the thermodynamics for Ag-Li mixing are favorable, the kinetics at the interface serve as an impediment. In battery cells, it has been shown that at larger rates of charging (larger than 4 mA cm$^{-2}$), Ag persists through charging, the high Li-content phases do not appear, and dendrites form in the battery \cite{magdysyuk2023}. Using cyclic voltammetry and CALPHAD, Corsi et al.\ found that while a nanostructured Ag anode would form a Li-rich solid solution, a bulk Ag anode absorbs a limited amount of Li, leading to Ag-Li intermetallics only forming on the surface. The bulk of the Ag anode remains as near-pure Ag\cite{corsi2022impact}. If the diffusion distances are short, like in Ag nanoparticles, the Li rapidly dissolves into Ag and forms Li-rich intermetallics. Thomas et al.\ calculated migration barriers for Li diffusion in the Ag-Li system and found that the D0$_3$-\ce{AgLi3}, $\gamma_1$, and $\gamma_3$ intermetallics all have very small migration barriers (0.1 eV), leading to rapid Li migration \cite{thomas2024thermodynamic} once these intermetallics form. These results suggest that a more thorough understanding of the interface is necessary for a complete picture of Ag-Li alloying in batteries.

While previous work has focused on the ground state Ag-Li structures, the interface can lead to substrate-induced phase transitions. For example, molecular dynamics simulations of Li solidification on a Li$_2$O substrate has suggested that amorphous Li forms at the Li-Li$_2$O interface. This amorphous region is characterized by disordered HCP-like pockets that show local configurations of HCP or FCC-like stacking and has a 5-layer thickness before a nearly-pure BCC Li phase is formed \cite{yang2023lithium}. Other material systems show similar behavior: epitaxial stabilization of four different polymorphs of VO$_2$ through in-plane lattice matching between the substrate and the desired structure \cite{lee2016epitaxial}, monoclinic to rhombahedral phase transition in La$_2$NiMnO$_6$ \cite{wu2018b}, tetragonal to orthorhombic phase transition in SrTiO$_3$ \cite{pertsev2000phase}, suppression of phase separation in FeSe$_{1-x}$Te$_x$ by plating on CaF$_2$ substrates \cite{imai2017control}, or the stabilization of the 1T-TaS$_2$ commersurate charge-density wave state using strain \cite{svetin2014transitions}. When considering the Ag/Li interface, we find the DFT formation energies per atom (using BCC Li as the reference state) are $-0.001$ eV/atom and 0.008 eV/atom for FCC and HCP Li respectively. This agrees with DFT calculations by Jerabek et al.\ who found that the FCC and BCC phases are quasi-degenerate \cite{jerabek2022solving}. Zero-point energies, which we did not include in our calculations, are necessary to stabilize the BCC Li phase compared to the FCC Li phase. With the zero-point energies, the BCC phase becomes 0.0033 eV/atom lower in energy than the FCC phase. This is because the 12 neighbors from the FCC structure (compared to the 8 neighbors for BCC) leads to larger zero-point energies. Additionally, the energy barrier for the Bain (BCC to FCC) or Burgers path (BCC to HCP) have barriers that are at most 1--2 meV/atom \cite{jerabek2022solving, thomas2024thermodynamic}. This means that Li can almost freely transform between the three phases.

In this work, we determine the atomic structure of Ag and Li interfaces and calculate vacancy formation energy and migration barriers through the interface to explore the onset of alloying between Ag and Li. First, we benchmark the pre-trained universal MACE-MP-0 potential against surface and defect energies for bulk Ag and Li against DFT values. Using a machine-learned potential will allow us to access the larger system sizes needed to model semicoherent interfaces. Second, we construct interfaces between FCC Ag and both BCC and FCC Li to determine the lowest energy interfacial structure. We include the BCC and FCC crystal structures for Li because the bulk formation energies are near degenerate. Third, we calculate vacancy formation energies and migration energies for migration through the interface to model the onset of alloying between Ag and Li show that small, almost-zero migration barriers for vacancies to move from the Ag to the Li promote mixing.

\section{Results and Discussion}

\begin{table}[htb]
\centering
\caption{Energy barriers for vacancy exchange calculated with DFT and MACE-MP-0. We calculate both vacancy-solvent and vacancy-solute exchange for both Ag and Li and find that the pure vacancy migration barriers are underestimated by MACE likely due to potential softening, but that the vacancy-solute exchange barriers are larger for MACE.}
\begin{tabular}{llcc}
\hline\hline
&&\multicolumn{2}{c}{Migration Barrier [eV]}\\
Solvent & Solute & DFT  & MACE-MP-0\\ 
\hline 
Ag & v$_\text{Ag}$ & 0.57 & 0.52 \\
Ag & Li$_\text{Ag}$ & 0.21 & 0.25 \\
Li & v$_\text{Li}$ & 0.038 & 0.017 \\
Li & Ag$_\text{Li}$ & 0.14 & 0.17 \\ 
\hline\hline
\end{tabular}
\label{tab:vacancy_agli}
\end{table} 

To benchmark the MACE-MP-0 potential, we find that the vacancy formation energy, migration barriers, and surface energies are within 10\% of the DFT values as shown in \Tabs{vacancy_agli}{surfaces}. The vacancy formation energy and migration energies calculated with MACE for bulk Ag and Li are within 0.05 eV of the DFT-calculated values. For Ag, the vacancy formation energy is 0.74 eV with MACE, which is larger than the 0.70 eV calculated with DFT. On the other hand, for Li vacancies, the formation energy was 0.57 eV with the MACE potential compared to 0.52 eV for DFT. This is within the estimated error for the MACE pre-trained potential, as vacancy formation energies can be underestimated by up to 20\% compared to the DFT-calculated values in bulk metals. For the exchange energies, we calculate the vacancy-solvent and vacancy-solute exchange barriers and find that the maximum deviation between MACE-MP-0 and DFT calculations was also 0.05 eV, with the vacancy-solvent barriers being lower with MACE-MP-0. An estimate of migration barriers errors from a systematic study of Mg ion migration by Deng et al.\cite{deng2025systematic} found a MAE of 0.34 eV; our comparison here is below that error estimate.
For low index surfaces (100), (110), and (111), we compare the surface energy $\gamma$
\be \gamma = \frac{E_\text{slab} - N\cdot E_\text{bulk}}{2A} \ee
with energy of the slab $E_\text{slab}$, bulk energy $E_\text{bulk}$, for $N$ atoms and area $A$ of the slab. We find that MACE-MP-0 surface energies deviate from the DFT values of Tran et al.\cite{tran2016surface} by at most 9\% in \Tab{surfaces}. Moreover, the relative order of the energies is the same for the three surfaces observed in our work. Thus, we conclude that the MACE-MP-0 potential can accurately represent the DFT energetics for our Ag-Li potential.

\begin{table}[htb]
\centering
\caption{Surface energies calculated with DFT and MACE-MP-0 for the (100), (110), and (111) surfaces. The DFT-calculated values are reproduced from Tran et al.\cite{tran2016surface}. For our system, we find that MACE agrees with DFT over the relative energetic order of these three low-index surfaces and that MACE accurately predicts the surface energies within 10\%.}
\begin{tabular}{llcc}
\hline\hline
&&\multicolumn{2}{c}{Surface Energy [eV/\AA$^2$]}\\
Bulk & & DFT & MACE-MP-0 \cite{tran2016surface}\\ 
\hline 
Ag & (100) & 0.818 & 0.806 \\
& (110) & 0.866 & 0.844 \\
 & (111) & 0.764 & 0.698 \\
Li & (100) & 0.462 & 0.482 \\ 
& (110) & 0.501 & 0.494 \\
 & (111) & 0.544 & 0.583 \\
\hline\hline
\end{tabular}
\label{tab:surfaces}

\end{table}

\begin{figure*}[htbp]
\centering\includegraphics[width=\wholefigwidth]{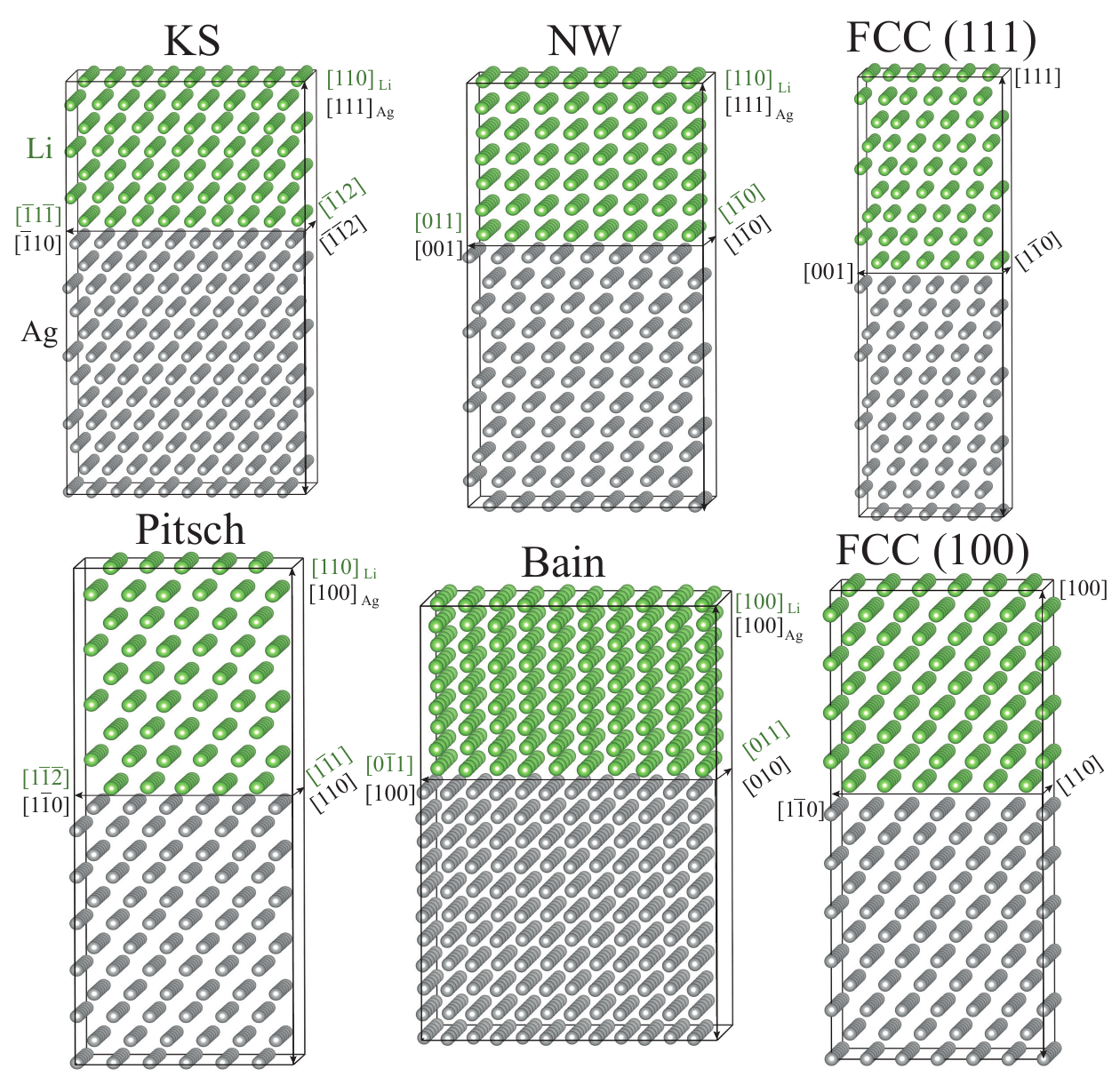}
 \caption{Schematic showing the unrelaxed interface orientations: KS, NW, FCC (111), Pitsch, Bain, and FCC (100). The KS interface differs from the NW orientation by a rotation of 5.26$^{\circ}$, and involves the close-packed plane for both FCC Ag and BCC Li. For the FCC (111) and FCC (100) interface orientations, we align the surface directions, and there is no relative rotation. For all six interfaces, all the strain is present in the Li surface, and the Ag has its relaxed bulk lattice constant (4.17 \AA).}
 \label{fig:geometry_ks}
\end{figure*}

We construct six different interface orientations between FCC Ag and Li as shown in \Fig{geometry_ks}. With BCC Li, we construct the Kurdjumov-Sachs \cite{kurdjumow1930mechanismus}, Nishiyama-Wasserman \cite{nishiyama1934x, wassermann1935ueber}, Pitsch \cite{pitsch1959martensite}, and Bain \cite{bain1924nature} interface orientations. 
With FCC Li, we construct two interface orientations: (111)$_{\mathrm{Ag}} \parallel$ (111)$_{\mathrm{Li}}$ and (100)$_{\mathrm{Ag}} \parallel$ (100)$_{\mathrm{Li}}$, which we will refer to as FCC (111) and FCC (100) respectively. The two FCC interfaces are constructed so that the surface directions are aligned, and there is no relative rotation. For the two FCC Li interfaces, we construct coherent interfaces with which 3.77\% compression of the Li for both in-plane directions. Because the lattice mismatch between Ag (4.17 \AA) and BCC Li (3.5 \AA) is $-16$\%, we construct semi-coherent interfaces and introduce misfit edge dislocations. The dimensions of the semi-coherent interfaces were chosen so that the so that the lattice mismatch in any direction was less than 4\%. For three of the four BCC Li interfaces (KS, NW, and Pitsch), we only need a mismatched number of unit cells for one of the in-plane directions. For the Ag slabs, the thickness is 24 \AA, and we fix all atoms within 12 \AA\ of the free surface to mimic bulk. The 12 \AA\ distance is chosen because it is the interaction cutoff for the used MACE-MP-0 potential. The thickness of the Li slab is chosen so that the interfacial energy is converged to within $0.01$ eV/\AA$^2$. To find the lowest energy stacking configuration, we performed a recursive grid search over the surface. For each step, we covered the area with 125 equally spaced translations with 5 in each direction, including the out-of-plane interfacial separation, and relaxed the structures using MACE-MP-0. The next step in the grid search uses a grid centered at the lowest energy structure, which are bounded in each dimension by translations that are one step away.

We relax the optimized final structures and find that Li preferentially plates in a FCC structure on Ag, but due to small energy differences, the KS or Pitsch interface orientations are still possible. The work of adhesion $W_\text{ad}$ shown in \Tab{workofadhesion} measures the amount of work needed to separate the two surfaces, and is defined as
\be W_\text{ad} = \frac{E_{\mathrm{Ag}} + E_{\mathrm{Li}} - E_{\mathrm{int}}}{A} , \ee
where $E_{\mathrm{Ag}}$ is the energy of the Ag slab, $E_{\mathrm{Li}}$ is the energy of the Li slab, $E_{\mathrm{int}}$ is the energy of the interface, and $A$ is the interfacial area.
Seymour and Aguadero used DFT to study coherent interfaces between BCC (111) Li and the following substrates: LiCl, Li$_3$OCl, LiMg, Li$_2$O, $\gamma$-Li$_3$PO$_4$, and AlSc. Our values for the $W_\text{ad}$ are similar to their metallic substrates (LiMg and AlSc), which had $W_\text{ad}$ of 0.08301 and 0.10242 eV/\AA$^2$.
Among all six interfaces studied in this work, the FCC (111) Li interface has the highest work of adhesion and is thus, the most energetically favorable. This is likely  because FCC Li allows for the formation of coherent interfaces and less lattice mismatch. 
The small $W_\text{ad}$ difference between the KS and FCC (111) and between the Pitsch and FCC (100) make it impossible to confidentially conclude that the FCC phase is favorable. The small differences are well within the error between MACE-MP-0 and DFT, and the slight preference for the FCC phase may be due to inaccuracies in the used potential.
For the two coherent interfaces, we also calculate the $W_\text{ad}$ using DFT and obtained a $W_\text{ad}$ of 0.093 eV/\AA$^2$ for FCC (111) and 0.106 eV/\AA$^2$ for FCC (100) showing that the interfaces are more stable than predicted by MACE. This is due to energetic differences between MACE and DFT, with the largest error for the Ag slabs. These Ag slab differences were on the order of 0.44 eV/atom.
Thus, a more careful examination of the relaxed structures must be completed before we can conclude that the FCC phase is preferred.

\begin{table}[htb]
\centering
\caption{Work of adhesion for the 6 interfaces calculated using MACE-MP-0 studied in this work: KS, NW, FCC (111), Pitsch, Bain, and FCC (100). For both the (111) and (100) Ag surfaces, the interfaces with FCC Li are energetically favorable. The FCC (111) is the most stable interface, followed by the KS.}
\begin{tabular}{lccc}
\hline\hline
Interface & Ag  & Li  & $W_\text{ad}$ [eV/\AA$^2$] (MACE) \\ \hline 
KS & (111) & $(110)_\text{BCC}$ & 0.0916 \\
NW & (111) & $(110)_\text{BCC}$  & 0.0898 \\
FCC (111) & (111) & $(111)_\text{FCC}$  & 0.0919 \\
Pitsch & (100) & $(110)_\text{BCC}$ & 0.0455 \\
Bain  & (100) & $(100)_\text{BCC}$ & 0.0170 \\
FCC (100) & (100) & $(100)_\text{FCC}$  & 0.0455 \\
\hline \hline
\end{tabular}
\label{tab:workofadhesion}
\end{table} 

There is significant interfacial reconstruction for the BCC interfaces, and both the Bain, Pitsch, and NW interfaces undergo a transformation to a more close-packed structure. The KS interface, which has the lowest energy of all BCC interfaces, undergoes a shearing in the layers closest to the interface, but still retains its original BCC structure at the free surface. While it does not fully phase transform, the relaxed KS interface becomes partially HCP-like.
The (110) plane becomes the (0001) plane in HCP, and the [111] direction transforms to [11$\overline{2}$0]. This is similar to the Burgers orientation relationship \cite{burgers1934process}. Since HCP Li is less energetically favorable than FCC Li, we believe that the phase transformation to HCP instead of FCC is because the structure retains its original ABA stacking from the BCC structure, not because HCP Li is energetically preferable. 
To match the underlying Ag lattice, we apply compressive strain in the [$\overline{1}$12] direction to the Li which is 27$^\circ$ from the direction that compressive strain is applied for the Burgers path. We think that this applied strain direction favors the Burgers transformation over the Bain path to FCC.
Additionally, the stacking fault energy in FCC and HCP Li is low, so much so that it was previously suggested that the ground state structure of Li was the 9R structure. The 9R structure is a close-packed lattice with multiple stacking faults (ABC\textbf{B}CA\textbf{C}AB\textbf{A}) \cite{berliner1986effect, zdetsis1986crystal}, and thus, the stacking faults, which separate HCP from FCC, are low in energy. 
Ackland et al.\cite{ackland2017quantum} cooled Li through molecular dynamics with an interatomic potential to find that the the BCC to FCC transition was kinetically hindered and only occurs when all the close-packed stackings are favorable compared to BCC.
The Pitsch interface similarly also undergoes a phase transition from the original (110) BCC Li surface to (100) FCC Li and forms a coherent interface, which has the same interfacial structure as the FCC (100) interface. This leads to the Pitsch having the same $W_\text{ad}$ as the BCC (100) phase. The Bain interface also undergoes a phase transformation, and forms the same interfacial structure as the FCC (100) interface. The extra Li atoms in the interfacial layer are ejected into the subsequent layers, leading to large amounts of atomic disorder away from the interface. 
Thus, even though there may be errors in the calculated $W_\text{ad}$ due to using MACE, we are confident that the preferred crystal structure for Li deposition on Ag is FCC because even the structures that were explicitly constructed in the BCC phase underwent phase transformations into close-packed structures.

\begin{figure*}[htbp]
 \centering\includegraphics[width=\halffigwidth]{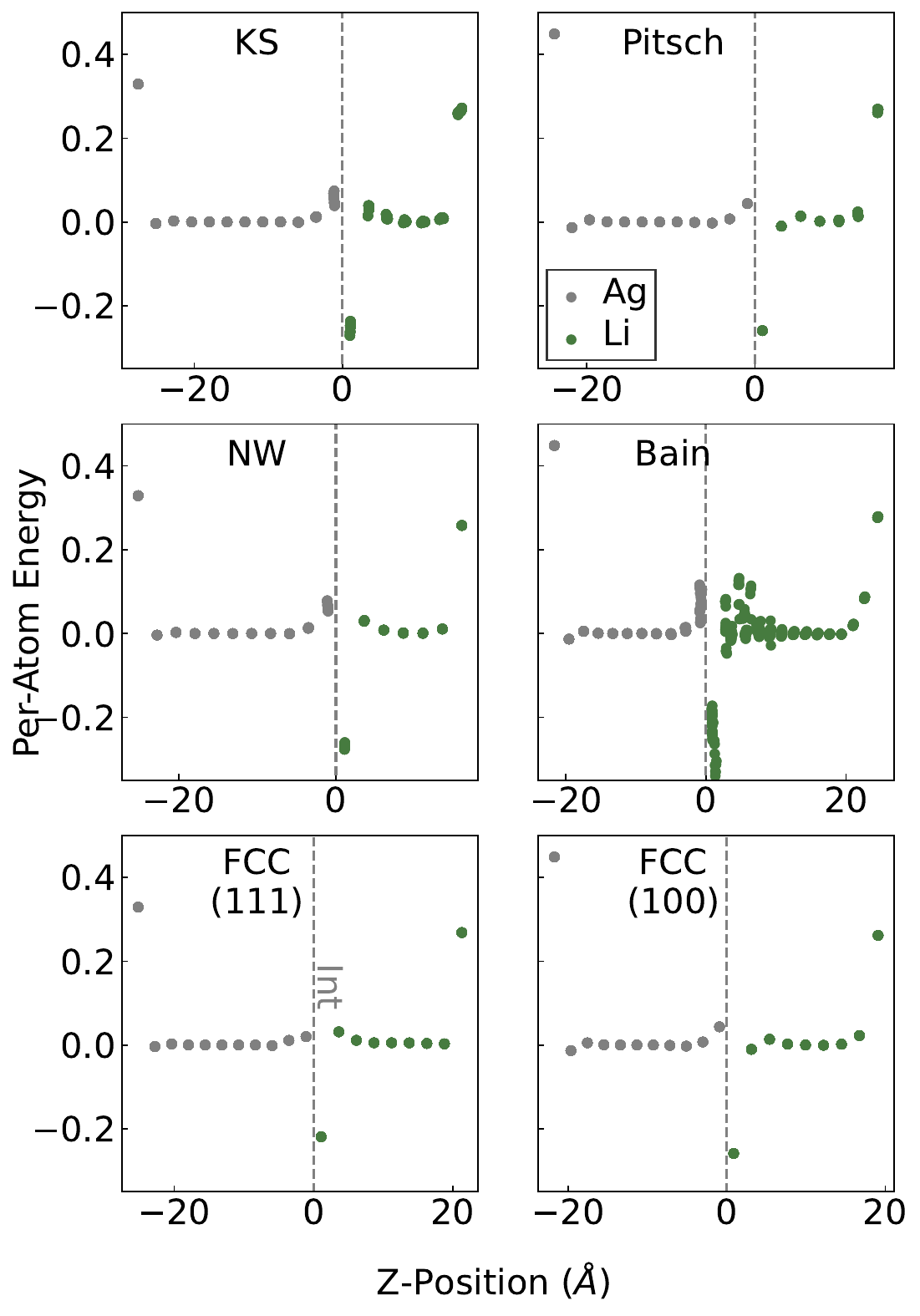}
 \caption{Difference between the per atom energies and the chemical potential calculated using MACE-MP-0. For all four interface orientations, we find that the first Li layer has smaller per atom energies, and larger per-atom energies for the first Ag layer respectively. The Bain interface has the most variation in the per-atom energies of the same layer due to increased atomic disorder after relaxation. The Pitsch interface has the least variation because a coherent interface is created after relaxation.}
 \label{fig:peratom}
\end{figure*}

The per-atom energies in \Fig{peratom} show that the first Li layer has 0.2 eV smaller per-atom energies compared to bulk, while the first Ag layer has larger per-atom energies. This suggests that the vacancy formation energy for the first Li layer should be larger than bulk, as the Li atoms are stabilized at the interface. This may be the reason why Li deposits smoothly on the Ag surface, as it is energetically favorable to be on the Ag surface. On the other hand, the Ag atoms at the interface have larger per-atom energies, and thus, leads to smaller vacancy formation energies than bulk. This may promote the migration of Ag vacancies towards the interface, which is a necessary step in alloying between Ag and Li.
For the Bain interface, there is the most variation in the per-atom energy due to the atomic disorder that occurs after the extra Li atoms in the interfacial layer is ejected into the Li bulk. Small variations [what is the numerical range] between atoms in the first Ag and Li layer for the semicoherent interfaces (KS and NW) due to the differences in the local structure. The two coherent interfaces (FCC (111), FCC (100)) have identical per-atom energies for atoms in the same layer. This is also true for the Pitsch interface, which phase transforms to be identical to the FCC (100) interface. The large per-atom energies for two free surfaces are unphysical and should be ignored.

\begin{figure*}[htbp]
 \centering\includegraphics[width=\halffigwidth]{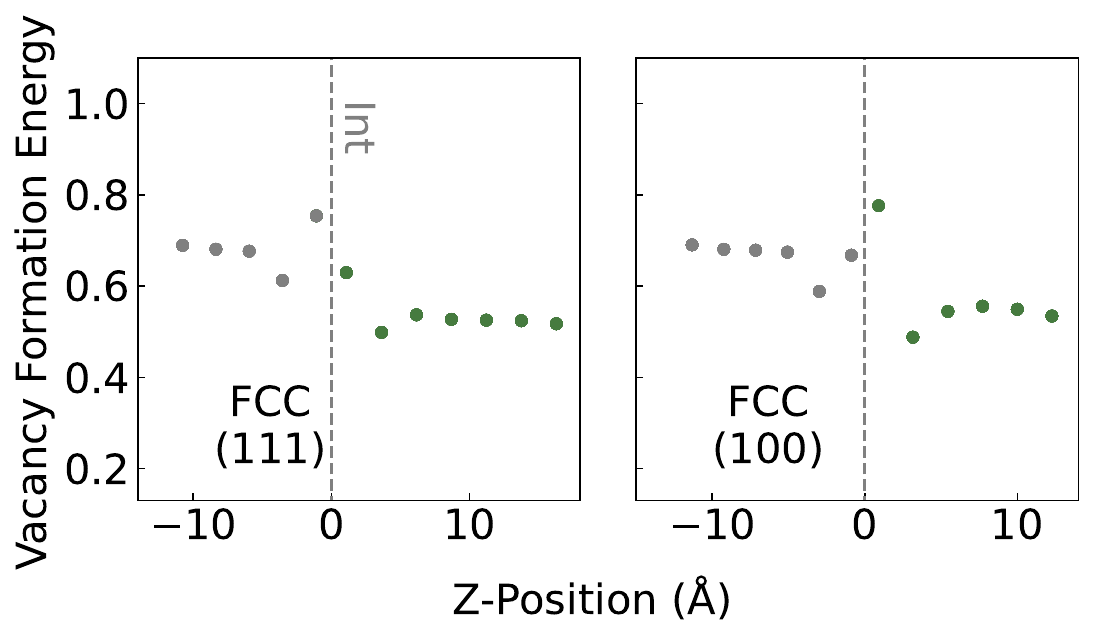}
 \caption{Vacancy formation energies calculated with MACE-MP-0 for the two FCC Li ORs. Vacancy formation energies were not calculated for fixed Ag atoms or Li atoms at the interface. Some Li atoms within 3 layers of the interface diffuse to the free surface and thus have non-physical vacancy formation energies were not included. Similarly, for the 12 \AA\,of Ag that we have fixed to mimic bulk, we did not calculate the vacancy formation energies.}
 \label{fig:vac_formation_ks}
\end{figure*}

In \Fig{vac_formation_ks}, we find that the vacancy formation energies increase at the interface for the two FCC interfaces, so there exists a driving force for vacancies to diffuse away from the interface into the Li bulk. As suggested by the per-atom energies, we see larger vacancy formation energies compared to the bulk vacancy formation energies for the layer of Li atoms closest to the interface.
 In the FCC (111) interface, the Li vacancy formation energy is 0.63 eV, 102 meV larger than the bulk value. Similarly, for the FCC (100) interface, the Li vacancy formation energy is 227 meV larger at the interface compared to bulk FCC Li.
For the lower energy FCC (111) interface, the Ag vacancy formation energy is 73 meV larger than the bulk Ag value for the Ag layer closest to the interface, but this is not true for the higher energy FCC (100) interface for which the Ag vacancy formation energy at the interface is almost identical to the bulk value (smaller by 5 meV).
This suggests an energetic preference for vacancies to migrate away from the interface, enabling smooth deposition and eventual alloying. As a result, we do not believe that mixing of Ag and Li directly at the interface is a rate-limiting step.
However, for both interfaces, we calculate that the vacancy formation energies are smaller than bulk for the Ag and Li layers one layer away from the interface. This implies that, while vacancies do not aggregate directly at the interface, larger concentrations of vacancies may cluster close to, but not directly at the interface. Thus, we suggest that the rate-limiting step is instead the migration of vacancies towards the interface from the sub-interface layers. Upon arrival, the vacancies can enable rapid mixing (see below), and eventual alloying between the Ag and Li.
These vacancy formation energies return back to their bulk values after 2--3 atomic layers.

\begin{figure*}[htbp]
\centering\includegraphics[width=\halffigwidth]{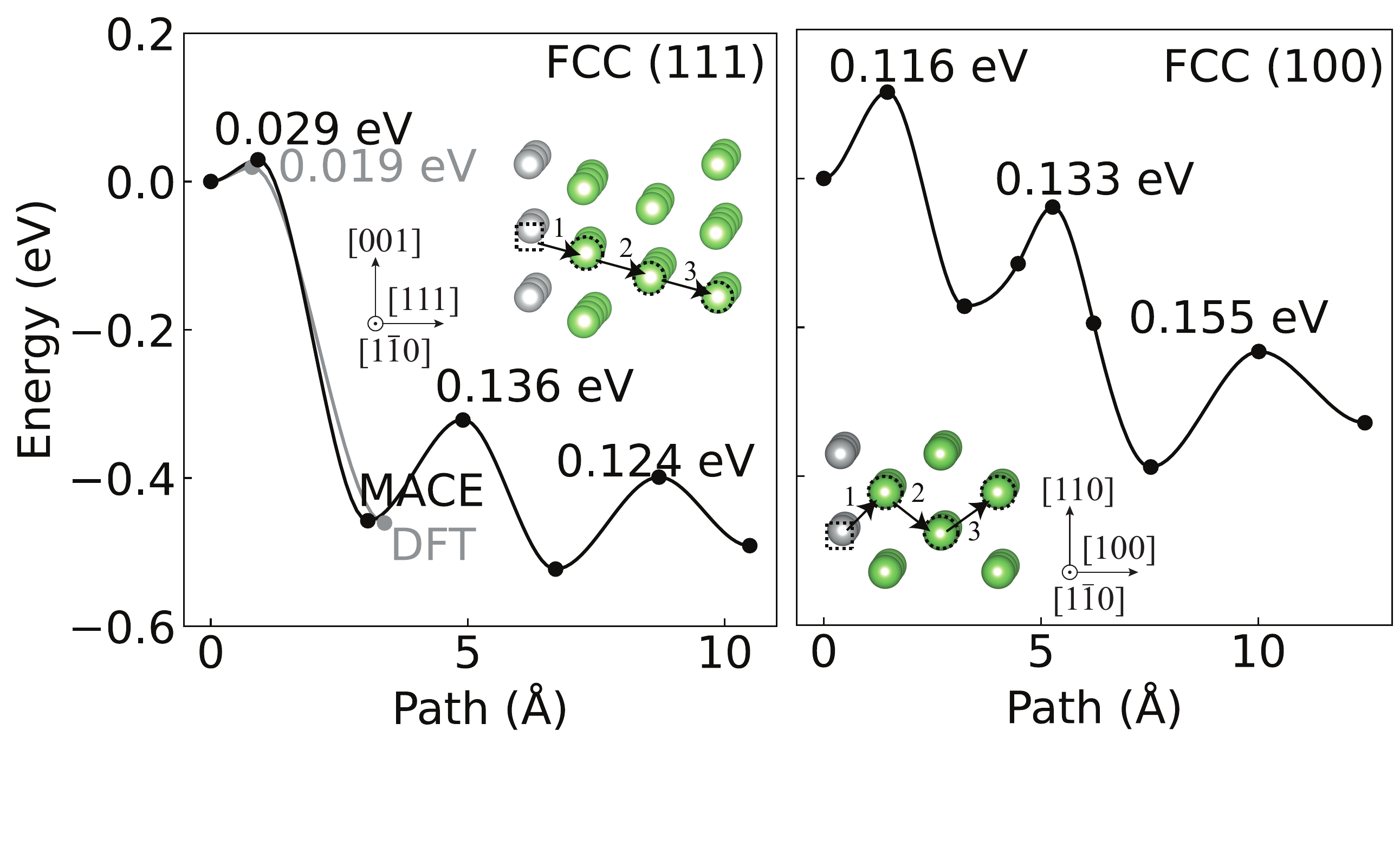}
 \caption{Migration barriers for vacancy migration in both the FCC (111) and FCC (100) interface. The black barriers are calculated with MACE-MP-0 while the gray one was calculated using DFT. While the migration barriers are very small for Li migration across the interface, the sub-interface migration barriers are significantly larger. For the (100) interface, the migration barriers are large even at the interface and are larger than the bulk Li migration barriers.}
\label{fig:neb}
\end{figure*}

In \Fig{neb}, we find small vacancy migration barriers across the Ag-Li interface, especially for the FCC (111) interfacial structure, showing energetically favorable mixing between Ag and Li. 
Both the MACE and DFT calculations (0.029 eV and 0.019 eV respectively) show almost negligible migration barriers for vacancies to migrate from the first Ag layer to the Li.
This supports our hypothesis that direct mixing between Ag and Li at the interface is rapid. Lithium will fill in any vacancies that occur on the Ag surface, leading to the formation of a perfect, void-free interface. 
This small migration barrier may explain why Ag reduces dendrite growth in batteries, as Li will spread out to cover the Ag surface and fill any gaps instead of gathering at hotspots.
However, the migration barriers for the vacancy to the 2nd and 3rd Li layer are larger than the bulk values (0.11 eV); thus, migration of vacancies through the Li away from the Ag-Li interface may kinetically hinder alloying. When multiple layers of pure Li form too quickly for alloying to occur, vacancies aggregate near the interface, voids may form in the sub-interface layers, and eventual dendrite formation may occur.
This is worsened by the slower kinetics in FCC Li compared to BCC Li. The vacancy migration barrier in FCC Li is larger than in BCC Li (35 meV \cite{behara2024fundamental}).
This may explain to the sluggish alloying behavior suggested by \cite{magdysyuk2023} who found limited alloying at high charging rates (larger than 4 mA cm$^{-2}$), upon which dendrites formed, leading to a catastrophic failure of their battery. Similarly, measurements done by \cite{corsi2022impact} have suggested that alloying between Ag and Li only occurs on the shell, with a pure Ag phase core. 

The less energetically-stable FCC (100) interface has much larger migration barriers shown in \Fig{neb}, which follows trends that smaller $W_\text{ad}$ also leads to trapping of vacancies at the interface \cite{yang2021maintaining}. Unlike the (111) interface, the migration barriers for vacancy migration from the Ag side of the interface to the Li are identical to the bulk FCC Li migration barriers (0.11 eV). Thus, there is a kinetic impediment for vacancy migration, which is not true for the more stable FCC (111) interface. Thus, in order to improve Li-Ag alloying, the less (100) surface present for Ag, the faster alloying should be.
Additionally, we once again see the same trend where the vacancy is energetically stable in the Li layers close to, but not at the interface, which will serve as a further impediment for alloying.

\section{Conclusions}

Due to the lattice compression of the FCC Li upon plating on Ag, it may be beneficial to increase the lattice constant of Ag through alloying. The only feasible possible candidate is Mg, as Mg is the only FCC or HCP element which forms a solid solution with both Ag and Li and increases the Ag lattice constant. From their phase diagrams, FCC Ag forms a solid solution with a solubility limit of 29.3 atomic \% of Mg \cite{nayeb1984ag}, while BCC Li forms a solid solution with Mg with up to 89.8 wt. \% Mg \cite{pelton1986ag}. Additionally, Mg has been shown experimentally to increase the Ag lattice constant \cite{letner1947x} from 4.06 \AA\ to 4.11 \AA\ at the solubility limit  \cite{letner1947x}.
(In situ?) Scanning electron microscopy found that Mg-Li alloys help split up voids and reduce void formation in batteries~\cite{zhao2025imaging}, so the presence of both Ag and Mg may aid in more uniform plating. Using DFT, we found that the substitutional energies for Mg in Ag, FCC Li, and BCC Li are $-1.16$ eV, 0.44 eV, and 0.56 eV/atom respectively, so Mg should remain alloyed with Ag over Li. This allows the Mg to act as a structural catalyst, without leaching from Ag during the charge-discharge cycle, which may further improve alloying kinetics between Ag and Li.

In this work, we conclude that Li preferentially plates on Ag as FCC Li to minimize lattice mismatch. While BCC interfaces were explicitly constructed, they underwent a phase transition into a close-packed structure. The (111) Ag surface is a more energetically favorable interface compared to the (100), which may result in both smooth Li deposition and faster alloying behavior. Due to low vacancy formation energies at the interface, and small migration barriers for vacancies to migrate across the interface from the Ag into the Li specifically for the (111) interface, there exists a driving force for vacancies to diffuse away from the interface into the Li bulk. It is likely that the initial alloying process between Ag and Li is rapid and not kinetically-hindered. These findings suggest possible alloying to impact the interface kinetics and reduce dendrite growth, which will increase the viability of solid state electrolyte batteries.

\section{Methods}

To perform our calculations, we use a combined DFT and machine-learned potential approach. The Vienna Ab initio Simulation Package (VASP) \cite{kresse1993ab, kresse1996efficiency, kresse1996efficient} was used for the DFT geometry optimization and vacancy formation energy calculations. We used the projector augmented wave (PAW) method with the Perdue–Burke–Ernzerhof (PBE) \cite{perdew1996generalized} formation of the generalized gradient approximation (GGA) of the exchange-correlation functional. We used a force-convergence of 0.005 eV/\AA, and a 500 eV plane-wave energy cutoff. A gamma-centered k-point grid mesh of $8 \times 8 \times 1$ was used with a Methfessel-Paxton \cite{monkhorst1976special} smearing width of 0.1 and a plane-wave energy cutoff of 500 eV. For our calculations with the MACE-MP-0 potential, we chose the medium MACE-MP-0 model, which has a maximum message equivariance $L=1$ as this provides a suitable cost-accuracy trade-off. We use the MACE-MP-0 potential \cite{batatia2023foundation} for relaxing our interfacial structures, calculating per-atom energies, and for calculating vacancy formation energies and migration barriers. For the smaller coherent interfaces, we also use DFT to verify our results.

\section*{Acknowledgements}
This work was supported by by the National Science Foundation Graduate Research Fellowship under Grant No. DGE-1746047 (G.L.). The data is available at the Materials Data Facility \cite{Blaszik2016, Blaszik2019}, doi:10.18126/gfpd-mm33 \cite{Lu2026Data}.

\printbibliography

\newpage

\sffamily Graphical Table of Contents:

\vspace{12pt}

\begin{minipage}[b][1.75in]{3.25in}
  \includegraphics[width=3.25in]{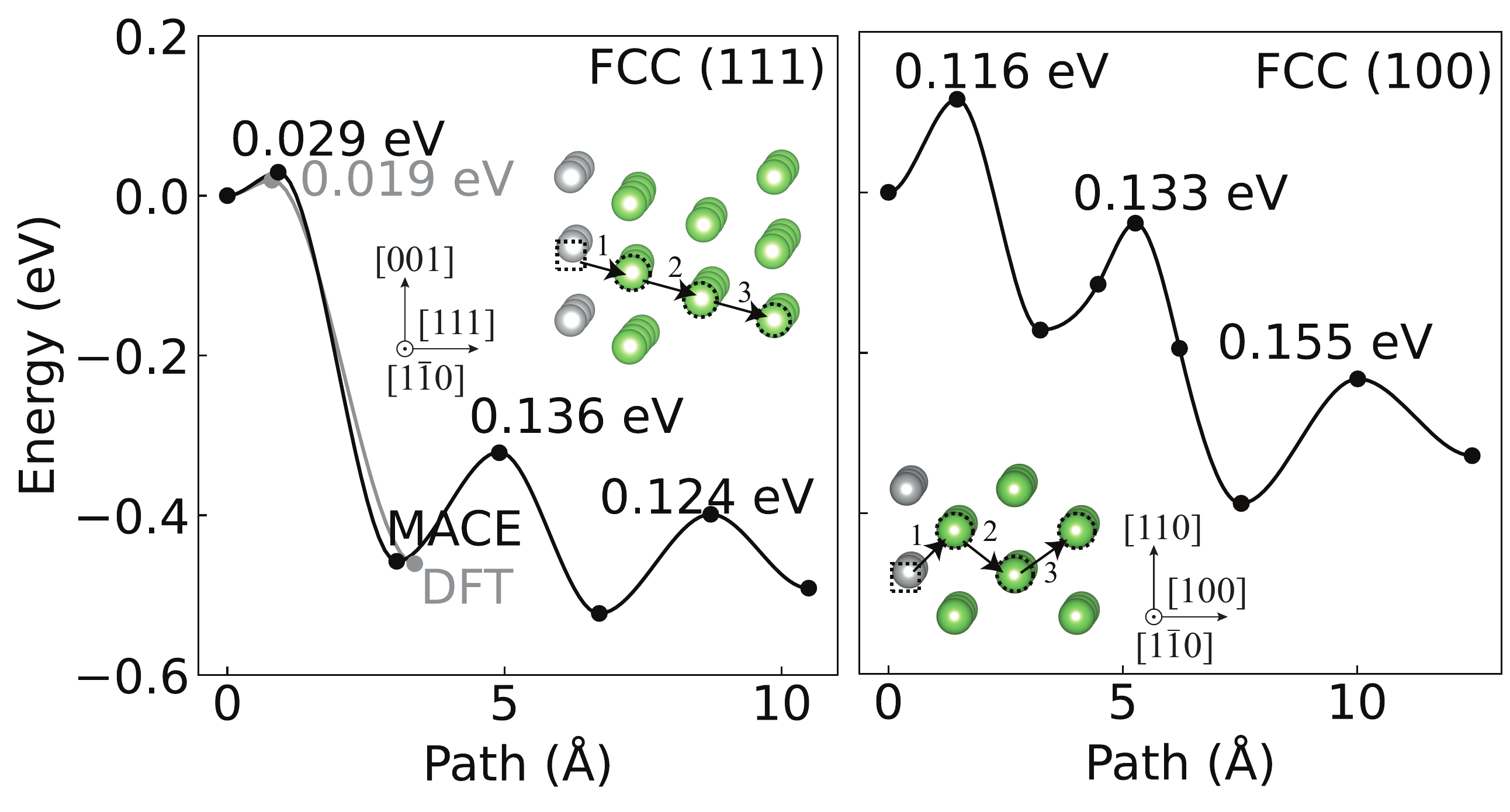}
  
\end{minipage}%

\end{document}